# Effects of Pressure on the Electronic Structures of LaOFeP

Yong Yang* and Xiao Hu

WPI Center for Materials Nanoarchitectonics (MANA)

National Institute for Materials Science, Tsukuba, 305-0044, Japan

We studied the electronic structures of LaOFeP under applied pressure using first-principles calculations. The electronic density of states at the Fermi level decreases continuously with increasing pressure. The electron branches of Fermi surfaces are rather robust to pressure, while the hole branches change significantly. Two hole surfaces shrink into small ellipsoid-like surfaces and disappear finally, at which the applied pressure is ~ 74.7 GPa. The pressure response can be understood by the band structures around the Fermi level. Comparative studies reveal that the disappearance of hole surfaces is mainly due to the compression of the FeP layer along the *c*-axis of unit cell.

---------------------------------------------------------------------------------------------------

*Present address: Global Research Center for Environment and Energy based on Nanomaterials Science (GREEN) at National Institute for Materials Science.

## I. INTRODUCTION

The Fe-based superconductors have received tremendous attention and interests ever since the discovery of superconductivity in the $LaO_{1-x}F_xFeAs$ series in the year 2008 [1]. Additional categories are synthesized and the reported superconducting transition temperature ($T_c$) increases rapidly from ~ 26 K to ~ 55 K [2-10]. Although the microscopic mechanism for the Fe-based superconductivity is far from clear, significant progresses have been achieved: For example, the stripe-like antiferromagnetic (AFM) phase of LaOFeAs at low temperature [11]; the BCS-like superconducting gap in $SmO_{0.85}F_{0.15}FeAs$ [12], and large isotope effects of Fe atoms in $SmO_{1-x}F_xFeAs$ and $Ba_{1-x}K_xFe_2As_2$ [13]; the co-existence of magnetism and superconductivity in $SmO_{1-x}F_xFeAs$ and $Ba_{1-x}K_xFe_2As_2$ [14, 15].

Basically there are two methods for inducing superconductivity: Charge doping and the application of pressure. Compared with charge doping, the application of pressure has the advantage of maintaining the stoichiometric composition of the superconducting materials and thus reduce the complexity of analysis. Up to now, pressure-induced superconductivity has been reported in LaOFeAs [16], $CaFe_2As_2$ [17] and $SrFe_2As_2$ [18]. First-principles calculations on LaOFeAs have shown that the electronic density of states (DOS) at the Fermi level of the AFM phase is significantly enhanced by applied pressure [19, 20]. Then, one may naturally ask, how about the pressure response of LaOFeP, which is the firstly reported Fe-based superconductor [21], and an analogue of LaOFeAs?

In this work, we studied the effects of pressure on the electronic structures of LaOFeP. Our first-principles calculations show that, the electronic DOS at the Fermi level decreases with increasing applied pressure. The hole pockets/cylinders of Fermi surfaces show significant changes under pressure. Two of three hole cylinders shrink into small ellipsoid-like surfaces and finally disappear with increasing pressure. This can be attributed to the up-shift of Fermi level by pressure. Compression along the *c*-axis of the unit cell is found to play a major role in the Fermi surface disappearance.

The content of this paper is organized as follows: after the introduction part, the modeling and computational details are described in Section II; the results are presented in Section III, including the pressure response of crystal structure and Fermi surface, the evolution of band structures and their orbital characteristics. This work is concluded in Section IV.

## II. COMPUTATIONAL METHODS

The calculations based on density functional theory (DFT) are performed by the Vienna *ab initio* simulation package (VASP) [22, 23], using a plane wave basis set and the PAW potentials [24, 25]. The exchange-correlation interactions are described by the generalized gradient approximation (GGA) with PBE type functional [26]. The energy cutoff for plane waves is 600 eV. A 12×12×6 Monkhorst-Pack k-points mesh [27] is used in the structural optimization and the calculation of electronic density of states (DOS). Our calculations begin with the experimentally determined crystal structure of LaOFeP [21, 28]. The tetragonal unit cell (Fig. 1(a)) of the non-magnetic (NM) phase is considered here. Although the ferromagnetic (FM) phase with weak magnetism (magnetic moment $M_{Fe} \sim 0.12$ $\mu_B$) is possible because of its slightly lower total energy (~ 2 meV/unit cell) with reference to the NM phase, no experimental evidence has been reported yet. Moreover, such a small energy difference is within the error bar of computation. In contrast to LaOFeAs [11], the experimental magnetic phase of LaOFeP is paramagnetic [21]. Moreover, the FM phase will dramatically switch to the NM phase upon volume compression ($\Delta V/V \sim 5\%$), due to the lower energy of the latter and the disappearance of magnetism under pressure. Therefore, choosing the NM phase as starting point has no effects on the conclusions of this work because the volume compression ratio here is well above 5%.

## III. RESULTS AND DISCUSSION

To study the effects of pressure, we gradually changed the volume of the crystal unit cell and relaxed the atomic positions and unit cell axes for each fixed volume. For the sake of clarity, the results are divided into two subsections. The evolution of crystal structure and electronic density of states (DOS) under pressure are presented in subsection A, and that of Fermi surface and the related band structures are given in subsection B.

### A. Crystal Structure and Electronic DOS

Figure 1(b) shows the tetrahedral bonding structure of LaO and FeP layers. There are six La-O-La angles in each LaO tetrahedron and only two inequivalent ones ($\alpha_1$, α₂ in Fig. 1(b)) because of crystal symmetry. Similar geometry is found for the FeP tetrahedron, where two inequivalent P-Fe-P angles ($\beta_1$, $\beta_2$) present. The calculated total energy as a function of volume of LaOFeP unit cell is shown in Fig. 1(c). The data are least-squares fitted into the Murnaghan equation of state [29, 30],

$$E_{tot}(V) = \frac{B_0 V}{B_0'}[\frac{(V_0/V)^{B_0'}}{B_0'-1}+1] + E_{tot}(V_0) - \frac{B_0 V_0}{B_0'-1} \qquad (1)$$

where $B_0$ and $B_0'$ are the bulk modulus and its pressure derivative at equilibrium volume $V_0$, at which the total energy of the system has its minimum. Compared to the experimental volume ($\Omega_0$) of LaOFeP at ambient pressure [21, 28], the volume $V_0$ obtained by our DFT calculation is slightly smaller ($V_0 = 0.98\Omega_0$, see Fig. 1(c)). The deduced values for $B_0$ and $B_0'$ are 103 GPa and 4.1, respectively. Compared with the bulk modulus of LaOFeAs, which is ~ 75 GPa [31], the one of LaOFeP is higher by ~ 37%. Thus, LaOFeP crystal is harder than LaOFeAs. Upon volume variation, the pressure exerted on the crystal is as follows:

$$P = -\frac{\partial E}{\partial V} = \frac{B_0}{B_0'}[(\frac{V_0}{V})^{B_0'} - 1] \qquad (2)$$

For direct comparison with experiments, the varying volume is expressed in units of $\Omega_0$. The pressure at volume $V = 1, 0.9, 0.8$ and $0.7\Omega_0$ is calculated to be -2 GPa, 10.5 GPa, 32.6 GPa, and 74.7 GPa, respectively. The negative pressure at $V = \Omega_0$ is caused by the small deviation of theoretical equilibrium volume ($V_0$) from the experimental value. For a given volume $V$, the error of calculated pressure resulted from the equilibrium volume deviation ($\Delta V_0$) is estimated to be $\Delta P = B_0[(\frac{V_0}{V})^{B_0'-1}](\frac{\Delta V_0}{V})$.

The structural parameters regarding the dimension of the unit cell and interatomic bonds of LaOFeP at $V = 1, 0.9, 0.8$, and $0.7\Omega_0$ are listed in Table I. With increasing pressure, the length of cell axes ($a$, $b$, $c$) and La-O and Fe-P bonds keep decreasing. Compared to $a$ (or $b$)-axis, larger magnitude of compression is found along the $c$-axis ($z$-direction in our case), which is the stacking direction of LaO and FeP layers. The pressure response of La-O-La angles and P-Fe-P angles are different: the La-O-La angles ($\alpha_1$, $\alpha_2$) show some fluctuation while the P-Fe-P angles ($\beta_1$, $\beta_2$) vary monotonically with volume contraction.

Figure 2(a) shows the calculated total electronic density of states (DOS) around the Fermi level for $V = 1, 0.9, 0.8$ and $0.7\Omega_0$. The height of the peak closest to the Fermi level decreases with contracted volume and the peak position keeps shifting up along the energy axis with decreasing volume. The electronic DOS at the Fermi level, $N(E_F)$, as a function of crystal volume is shown in Fig. 2(b). A general tendency of decreasing (increasing) $N(E_F)$ with decreasing (increasing) volume is found. Such pressure response of LaOFeP is distinct from that of LaOFeAs, whose $N(E_F)$ is enhanced within a certain range of applied pressure [19, 20]. Similar behavior is expected for the physical quantities that are positively correlated with $N(E_F)$, such as the electron contribution to specific heat, Pauli paramagnetic susceptibility, and the electrical conductivity. The partial DOS from the La, O, Fe, P atoms are shown in Figs. 2(c)-2(d), for the volume at $V = \Omega_0$ and $0.7\Omega_0$. It is clear that the Fe and P atoms contribute to the major part of the total DOS around the Fermi level, and the Fe atoms have the largest contribution. This implies the key role of the FeP layer in the onset of superconductivity.

## B. Fermi Surfaces and Band Structures

The Fermi surfaces at $V = 1$, 0.9, 0.8 and $0.7\Omega_0$ are displayed in Figs. 3 to 6. The Fermi surfaces calculated at $V = \Omega_0$ compare well with the previous work [32]. Overall, the most striking change in the Fermi surfaces upon pressure variation is the two sheets in the (a) and (b) panels of Figs. 3-5 disappear in Fig. 6. To understand such changes, we plot the band structures at some high-symmetry k-point lines of the Brillouin zone for the corresponding cell volumes from $V = \Omega_0$ to $0.7\Omega_0$ in Fig. 7.

The corresponding energy bands of the Fermi surface branches in Fig. 3 are marked by the letters **a**-**e** in Fig. 7(a) with the same alphabetic order as in Fig. 3. The three bands **a**, **b** and **c** run across the Fermi level around the Γ point, and the sets of all the crossing points form the 3D Fermi surface sheets shown in Figs. 3(a)-(c). The electronic states enclosed by the each sheet are unoccupied because their energies are above the Fermi level. Thus, the sheets in Figs. 3(a)-(c) are hole surfaces. On the other hand, the bands **d** and **e** go through the Fermi level around the M point, resulting in two additional branches of Fermi surface in Figs. 3(d)-(e). From band structures, the electronic states inside the sheets in Figs. 3(d)-(e) are occupied by electrons. Therefore, they are electron surfaces. The Fermi surfaces in Figs. 4 to 6 can be similarly classified.

Under applied pressure, the geometries of the electron cylinders around the M point are rather robust and remain almost unchanged. This can be understood from the energy bands. The dispersion and relative positions to the Fermi level of bands **d** and **e** in Figs. 7(a)-(d) are almost invariant with decreasing volumes. On the contrary, remarkable changes take place in the hole surfaces and the corresponding bands. When the crystal volume is compressed from $V = \Omega_0$ to $0.9\Omega_0$, the two separate hole pockets along the $\Gamma Z$ line are in touch and form a simply connected hole cylinder (Fig. 4(a)). This can be explained by the variation of band **a** in Figs. 7(a)-(b). At $V = \Omega_0$, the crossing points between band **a** and the Fermi level sit along the $\Gamma Z$ line. No crossing point is found along the $\Gamma X$ or $\Gamma M$ line. The situation changes at $V = 0.9\Omega_0$. The crossing points locate along the $\Gamma M$ line, no crossing point is found along the $\Gamma Z$ line. Compared with Fig. 7(a), the position of band **a** referenced to the Fermi level is shifted up in Fig. 7(b). This changes the positions of the crossing points and modifies the shape of Fermi surface.

From Figs. 4 to 6, the first two hole Fermi surfaces contract and disappear eventually. Again, such behavior can be understood from the band structures. The dimensions of the hole surfaces can be measured by the lengths of the line segments joining the crossing points and the $\Gamma$ point. As seen from Figs. 7(b) and 7(c), although the line segments along the $\Gamma X$ and $\Gamma M$ lines are almost unchanged, the bands **a** and **b** approaches the Fermi level at points in-between the $\Gamma Z$ line, leading to a smaller line segment than $0.5b_3$, with $b_3$ being the length of reciprocal lattice vector in the $\Gamma Z$ direction. With further compression from $V = 0.8\Omega_0$ to $0.7\Omega_0$, the hole pockets in Figs. 5(a) and 5(b) continue to shrink and finally disappear (Fig. 6). Such evolution is due to the up-shift of the relative position of Fermi level to bands **a** and **b** in Figs. 7(c) and 7(d). The Fermi level in Fig. 7(d) is right above bands **a** and **b** and there is no any crossing points between them.

Using the functionality of site- and *lm*-projection implemented in VASP code [33], we analyzed the wave function characteristics of bands **a-e** in Fig. 7. For each k-point, the *lm*-component of each band is obtained by projecting it onto spherical harmonics within spheres of a radius $R_c$ around each atom and then the results are summed over all the atoms. In our calculation, $R_c$ = 1.53, 1.28, 1.30, 1.50 Å is adopted for La, O, Fe, P, respectively. Since the most remarkable changes take place in the corresponding Fermi surfaces of bands **a-c** at varying volumes, we will focus on the characteristics of the three bands. The calculated *lm*-components of bands **a-c** at Γ point for volume $V$ = 1, 0.9, 0.8, and $0.7\Omega_0$ are listed in Table II. For bands around the Fermi level, the contribution from La atoms is negligible comparing to that of O, Fe and P atoms. Therefore, the *lm*-components of angular quantum numbers higher than 2 ($l > 2$, i.e., f orbital from La) are not shown in Table II. The common feature is that, the wave functions are localized with the major component from the *d*-orbitals ($d_{z^2}$, or $d_{yz}$, $d_{xz}$) of Fe atoms. From Table II, the sum of the *lm*-components of each band is close to but smaller than 1 (normalization of wave function), which implies that the *lm*-projection has captured the major part but still lacks some of the characteristics. The missing feature is due to the reasons: 1) The total volume occupied by the spheres centered at the atoms is ~ 70.4 % of the unit cell volume; 2) The basis set employed for projection is not complete. Calculated eigenenergies at Γ point show that, bands **b** and **c** are degenerate at $V = \Omega_0$ and $0.9\Omega_0$ while bands **a** and **b** are degenerate at $V = 0.8\Omega_0$ and $0.7\Omega_0$. Indeed, we can find rotation symmetry between the *lm*-components of every two degenerate bands at $V = \Omega_0$ to $0.7\Omega_0$: the wave function of the two bands is identical upon rotation by 90° in real space, from *x* direction to *y* direction. The two-fold degeneracy reflects the equivalence of *x* and *y* directions in the tetragonal unit cell of LaOFeP. Going into details, the *lm*-components of two degenerate bands consist of $p_x$, $p_y$, $d_{yz}$, and $d_{xz}$ orbitals with the other components being zero. The hybridization of these orbitals corresponds to π bonding. The non-degenerate band only consists of *s*, $p_z$, and $d_{z^2}$ orbitals, which has the characteristics of σ bonds. With increasing applied pressure, the *s* and *p* components are increased, which indicates the enhancement of orbital hybridization. Further analysis reveals that, for all the energy bands at Γ point, the two-fold degeneracy will appear as long as the *lm*-component from one of the orbitals $p_x$, $p_y$, $d_{yz}$, and $d_{xz}$ is nonzero. Again, this is a consequence of the rotation symmetry between the *x* and *y* directions. Variation of $R_c$ within a reasonable range (not differs too much from covalent or ionic radii) will only slightly modify the numbers in Table II but the order of contribution from each orbital as well as the symmetry is kept.

The above studies assume a homogeneous applied pressure, which can be realized by modern experimental instruments. However, the magnitude of compression along the unit cell vectors is not homogeneous: Compression along $\vec{a}$ and $\vec{b}$ is equivalent, while compression along $\vec{a}$ (or $\vec{b}$) and $\vec{c}$ is inequivalent. This is due to the tetragonal layered crystal structure of LaOFeP. Consequently, compression under external pressure can be divided into two directions: along *ab*-plane (constructed by vectors $\vec{a}\,\&\,\vec{b}$) and along *c*-axis ($\vec{c}$ direction). Similar to LaOFeAs [19], the *c*-axis is more compressible than the *ab*-plane.

To find out the role of the two compression directions in the disappearance of Fermi surfaces, we studied the compression along the *ab*-plane and the *c*-axis separately. The *c*-axis is kept at the value of ambient-pressure when studying the *ab*-plane compression and vice versa. For each direction, the magnitude of compression is chosen to be the same as at $V = 0.7\Omega_0$, at which two hole Fermi surfaces disappear. The dimension of unit cell at $V = \Omega_0$ may be expressed as $\{a_0, a_0, c_0\}$, with $a_0$, $a_0$, and $c_0$ being the length of the three lattice vectors, then it is $\{0.936a_0, 0.936a_0, 0.798c_0\}$ for $V = 0.7\Omega_0$. For the study of *ab*-plane compression, the unit cell dimension is $\{0.936a_0, 0.936a_0, c_0\}$, and it is $\{a_0, a_0, 0.798c_0\}$ for study of compression along *c*-axis only. The Fermi surfaces are displayed in the left panels of Fig. 8, for compression along *ab*-plane and *c*-axis, respectively. The corresponding band structures are shown in the right panels. For *ab*-plane compression, the hole surface corresponding to band **a** (Fig. 8(h)) disappears, due to the fact that band **a** is lying right below the Fermi level. The other hole and electron surfaces are remained (Figs. 8(a)-(d)). In the case of *c*-axis compression, the two hole surfaces of bands **a** and **b** disappear (Figs. 8(e)-(g)). This is again explained by the down-shift of both bands with reference to the Fermi level (Fig. 8(i)).

With presence of applied pressure, both the LaO and FeP layers are compressed along the *ab*-plane and *c*-axis. The evolution of the distance of La atoms to the underlying O-plane, and the P atoms to the underlying Fe-plane is shown in Figs. 9(a)-(b). In spite of some fluctuations, the heights of La and P atoms with reference to the corresponding O and Fe planes show a general tendency of decreasing with volume contraction. Since the FeP layer contributes to the major part of the DOS near the Fermi level, one may expect that the variation of the electronic structures in Fig. 7 should be mainly attributed to the modifications of bonding geometries of the FeP layer. To demonstrate this point, we performed calculations at volume $V = \Omega_0$ for two structures. In the first structure, the $z$-coordinates of La atoms are chosen to be the values at $V = 0.7\Omega_0$ while the other coordinates are kept at that of $V = \Omega_0$, where only the LaO layer is contracted in the $c$-axis. In the second structure, the $z$-coordinates of P atoms take the values at $V = 0.7\Omega_0$ and the other coordinates take the values at $V = \Omega_0$, which models the $c$-axis contraction of FeP layer. The calculated band structures are shown in Fig. 9(c) and Fig. 9(d), respectively. One sees that at least two bands run across the Fermi level in Fig. 9(c) while all the valence bands sit right below the Fermi level in Fig. 9(d). This points directly to the key role of FeP layer contraction. Therefore, the disappearance of two hole Fermi surfaces is caused by compression along both directions, but with the major contribution from the compression along the $c$-axis of unit cell, or more precisely, the compression of the FeP layer in the $c$-axis.

The contraction and disappearance of the two hole Fermi surfaces have significant effects on the electronic structures. This is evidenced from the continuous decreasing electronic DOS shown in Fig. 2. Suppose that the strength of electron-phonon coupling only changes slightly upon applied pressure, within the framework of BCS theory [34] the superconducting transition temperature $T_c$ would continue to decrease with increasing pressure. The evolution of Fermi surfaces can be measured via the quantum oscillation of transport properties in a magnetic field, for example, the de Haas-van Alphen (dHvA) effect. From the band structures shown in Fig. 7, the energy dispersion in the *k*-space is rather steady in spite of some minor changes. That means that the effective mass of electrons, $m^* = \hbar^2 / |\nabla_k^2 E(k)|$, can be taken as constant within the range of applied pressures studied in the present work. Thus, at a given magnetic field *B*, the cyclotron frequency measured in the dHvA effect, $\omega_c = \frac{eB}{m^* c}$, can be regarded as invariant with respect to a wide range of applied pressure (from 0 to 74.7 GPa). From a previous theoretical work which showed that the cyclotron resonance frequency of a short-range interacting electron gas in a uniform magnetic field is independent of the interaction [35], the electron momentum as well as the electron-electron interaction in LaOFeP may be expressed in the same form as given in Ref. [35].

## IV. Concluding Remarks

To summarize, the effects of pressure on the electronic structures of LaOFeP have been studied using first-principles calculations. With the presence of applied pressure, the electronic structures are largely modified, and the electronic DOS at the Fermi level decreases continuously with increasing pressure. The two types of Fermi surfaces exhibit completely different response to pressure. The electron branches are rather robust to volume contraction whereas the hole branches change significantly in shape. With increasing pressure, two of the hole surfaces collapse into tiny ones and finally disappear when the crystal volume is reduced to ~ 70% of the experimental value at ambient pressure. Evolution of the Fermi surfaces can be understood by the band structures. The shrinking and disappearance of the hole surfaces are explained by the up-shift of Fermi level with reference to the energy bands. The characteristics of the bands around the Fermi level are analyzed, which reflect the symmetry of the unit cell. Comparative studies demonstrate that, the disappearance of hole Fermi surfaces is mainly caused by compression along $c$-axis of the unit cell, which reduces the relative distance of the P atoms to the Fe-plane with which they bonded. The changes of electronic structures under applied pressure should have significant effects on response properties such as the electronic heat capacity, the electronic transport properties, and the magnetic susceptibility.

## ACKNOWLEDGMENTS

Calculations were performed on Numerical Materials Simulator (SGI Altix) of NIMS. The Fermi surfaces are drawn using the program XCrySDen [36-38]. This work was supported by WPI Initiative on Materials Nanoarchitectonics, MEXT, Japan.

TABLE I. Calculated unit cell parameters at volume $V = 1, 0.9, 0.8$, and $0.7\Omega_0$: length of cell axes ($a$, $b$, $c$), length of La-O and Fe-P bonds (La-O, Fe-P) and angles of La-O-La ($\alpha_1$, $\alpha_2$) and P-Fe-P ($\beta_1$, $\beta_2$), see Figs. 1 **(a)-(b)**.

| Volume ($\Omega_0$) | $a$ (Å) | $b$ (Å) | $c$ (Å) | La-O (Å) | Fe-P (Å) | $\alpha_1$ (°) | $\alpha_2$ (°) | $\beta_1$ (°) | $\beta_2$ (°) |
|---|---|---|---|---|---|---|---|---|---|
| 1 | 3.946 | 3.946 | 8.589 | 2.357 | 2.239 | 113.70 | 107.40 | 123.57 | 102.92 |
| 0.9 | 3.859 | 3.859 | 8.085 | 2.311 | 2.193 | 113.23 | 107.63 | 123.19 | 103.08 |
| 0.8 | 3.772 | 3.772 | 7.520 | 2.253 | 2.150 | 113.65 | 107.42 | 122.61 | 103.33 |
| 0.7 | 3.695 | 3.695 | 6.858 | 2.185 | 2.110 | 115.44 | 106.57 | 122.27 | 103.48 |

TABLE II. Calculated *lm*-component of the bands **a**, **b**, and **c** (Fig. 7), at varying volume $V = 1, 0.9, 0.8, 0.7$, in the units of $\Omega_0$.

| Volume/Band | | $s$ | $p_y$ | $p_z$ | $p_x$ | $d_{xy}$ | $d_{yz}$ | $d_{z^2}$ | $d_{xz}$ | $d_{x^2-y^2}$ |
|---|---|---|---|---|---|---|---|---|---|---|
| 1 | **a** | 0.061 | 0.000 | 0.065 | 0.000 | 0.000 | 0.000 | 0.662 | 0.000 | 0.000 |
|   | **b** | 0.000 | 0.049 | 0.000 | 0.041 | 0.000 | 0.416 | 0.000 | 0.350 | 0.000 |
|   | **c** | 0.000 | 0.041 | 0.000 | 0.049 | 0.000 | 0.350 | 0.000 | 0.416 | 0.000 |
| 0.9 | **a** | 0.064 | 0.000 | 0.075 | 0.000 | 0.000 | 0.000 | 0.679 | 0.000 | 0.000 |
|   | **b** | 0.000 | 0.079 | 0.000 | 0.027 | 0.000 | 0.560 | 0.000 | 0.190 | 0.000 |
|   | **c** | 0.000 | 0.027 | 0.000 | 0.079 | 0.000 | 0.190 | 0.000 | 0.560 | 0.000 |
| 0.8 | **a** | 0.000 | 0.032 | 0.000 | 0.108 | 0.000 | 0.164 | 0.000 | 0.556 | 0.000 |
|   | **b** | 0.000 | 0.108 | 0.000 | 0.032 | 0.000 | 0.556 | 0.000 | 0.164 | 0.000 |
|   | **c** | 0.068 | 0.000 | 0.092 | 0.000 | 0.000 | 0.000 | 0.699 | 0.000 | 0.000 |
| 0.7 | **a** | 0.000 | 0.090 | 0.000 | 0.120 | 0.000 | 0.281 | 0.000 | 0.376 | 0.000 |
|   | **b** | 0.000 | 0.120 | 0.000 | 0.090 | 0.000 | 0.376 | 0.000 | 0.281 | 0.000 |
|   | **c** | 0.076 | 0.000 | 0.131 | 0.000 | 0.000 | 0.000 | 0.698 | 0.000 | 0.000 |

**Figures & Captions**

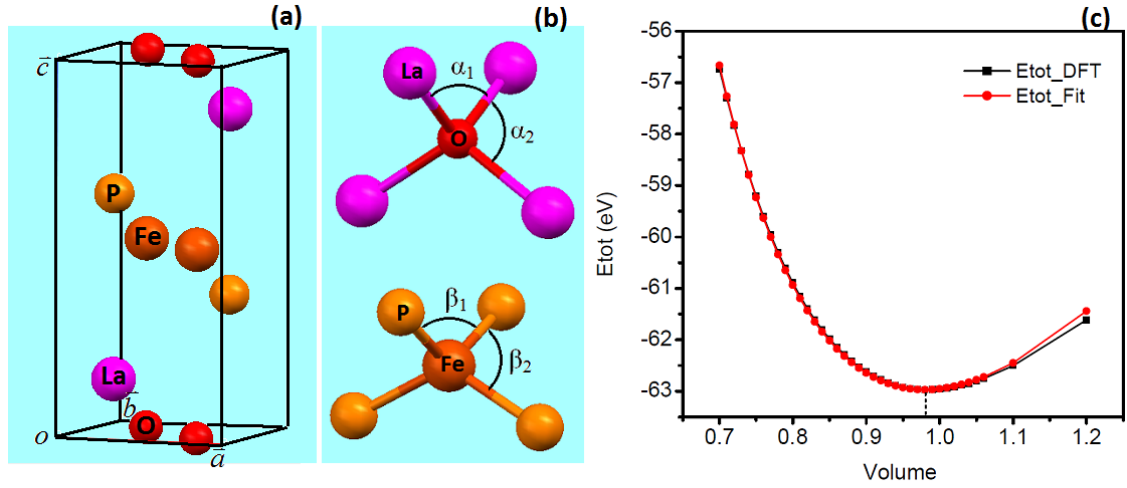

**FIG. 1** (color online) **(a)** Unit cell of the LaOFeP crystal. **(b)** Tetrahedrally bonded structures in LaO and FeP layers, with the La-O-La and P-Fe-P angles marked. **(c)** Total energy as a function of unit cell volume. The volume is normalized to $\Omega_0$, experimental equilibrium volume at ambient pressure. The theoretical minimum is indicated by a vertical dashed line segment. The solid curve joined by red dots is the fitting to Eq. (1).

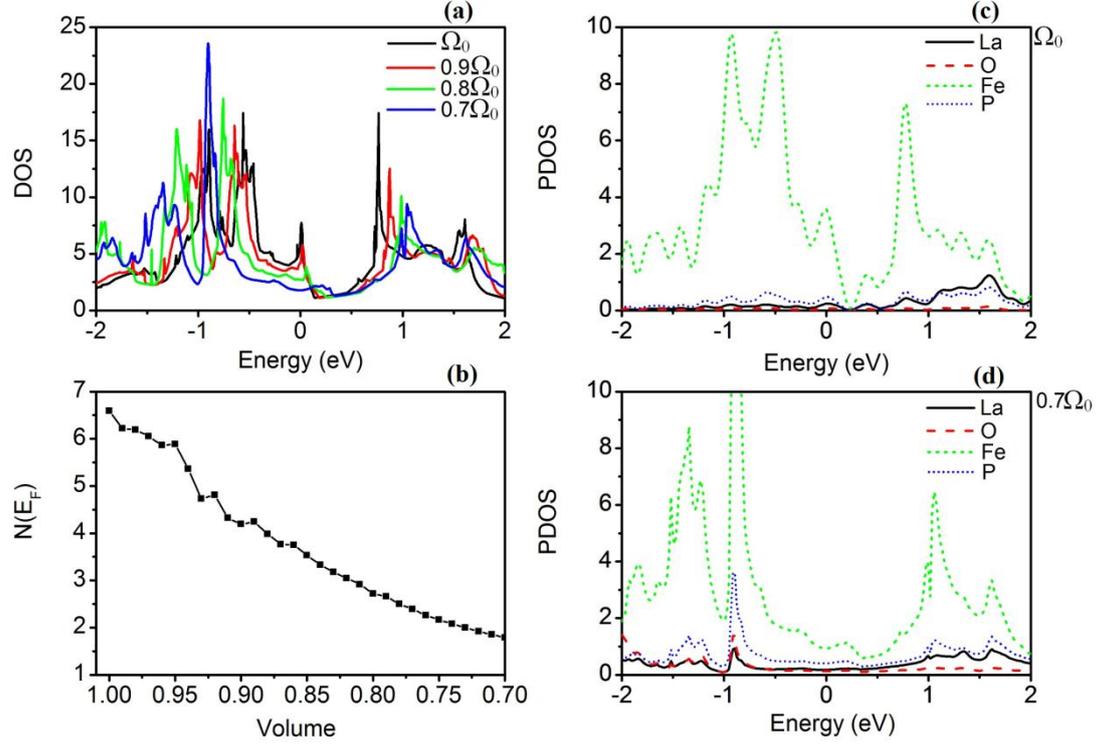

**FIG. 2** (color online) Left panels: **(a)** Calculated electronic DOS at $V = 1, 0.9, 0.8$, and $0.7\Omega_0$. The Fermi level is set at 0. **(b)** DOS at the Fermi level as a function of the normalized volume. Right panels: Partial electronic DOS (PDOS) at $V = \Omega_0$ (**(c)**) and $0.7\Omega_0$ (**(d)**). The data are in the units of states/eV/unit cell.

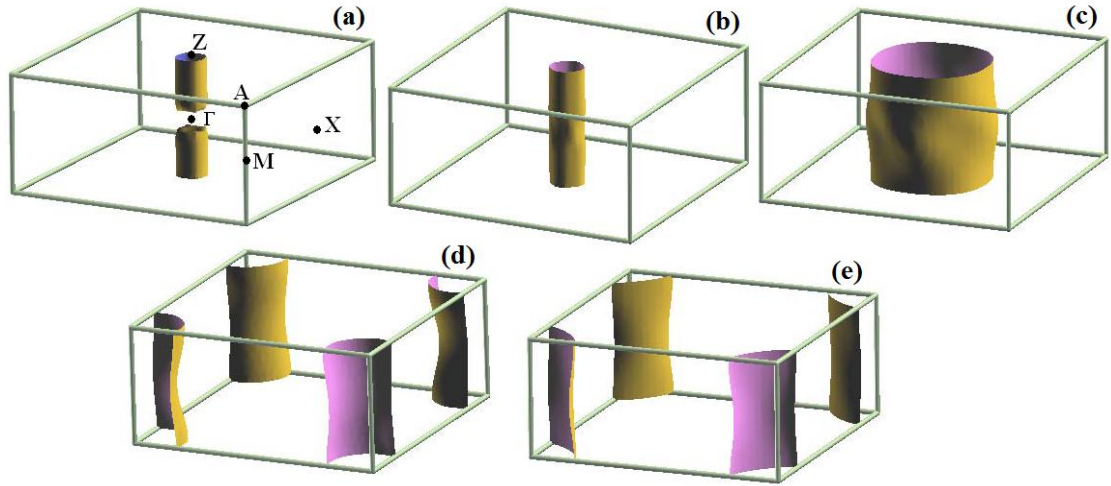

**FIG. 3** (color online) Fermi surfaces of LaOFeP at $V = \Omega_0$. Panels **(a)-(c)** are hole surfaces; panels **(d)-(e)** are electron surfaces. The fractional coordinates of the symmetry k-points in the Brillouin zone: $\Gamma = (0, 0, 0)$, $X = (0.5, 0, 0)$, $M = (0.5, 0.5, 0)$, $Z = (0, 0, 0.5)$, $A = (0.5, 0.5, 0.5)$. The convention of symmetry points applies to all the figures.

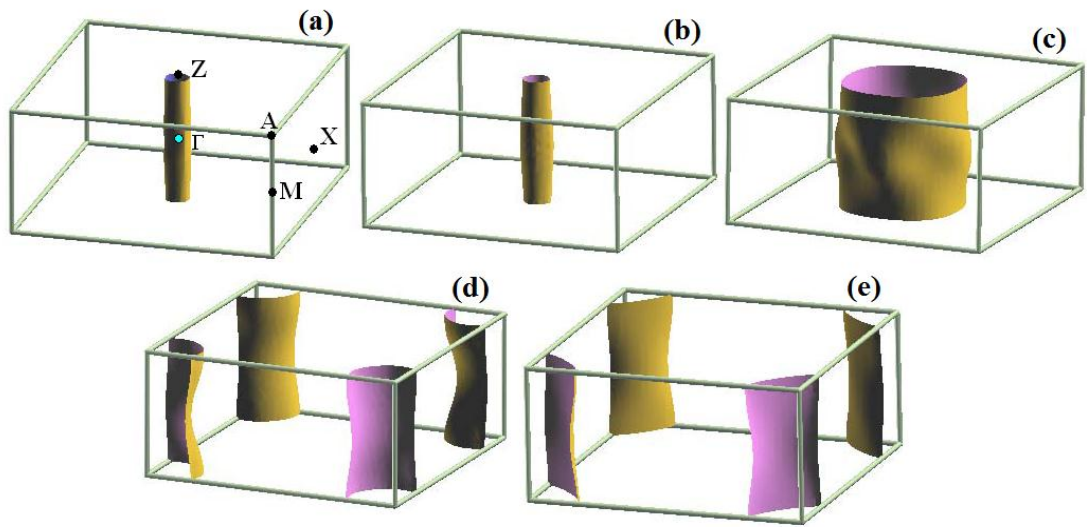

**FIG. 4** (color online) Same as Fig. 3 but for $V = 0.9\Omega_0$.

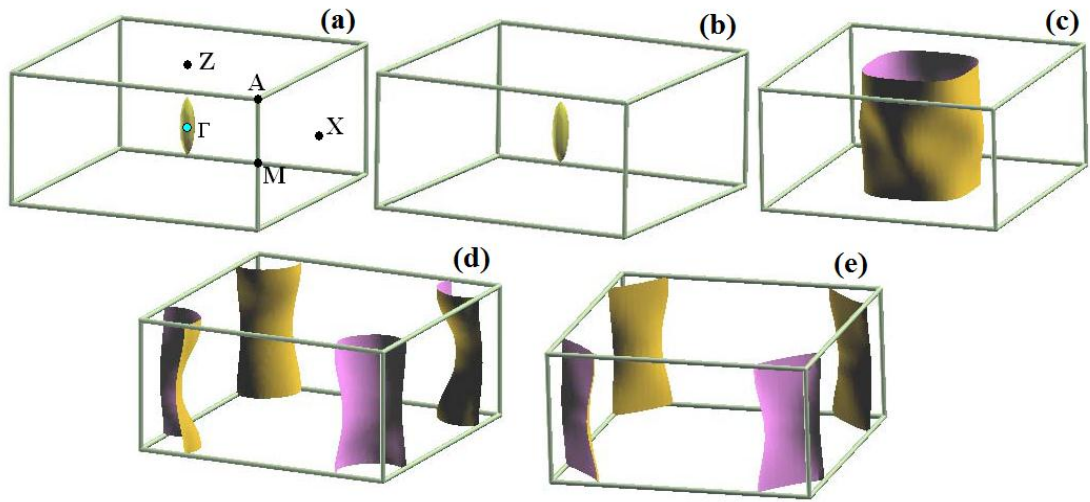

**FIG. 5** (color online) Same as Fig. 3 but for $V = 0.8\Omega_0$.

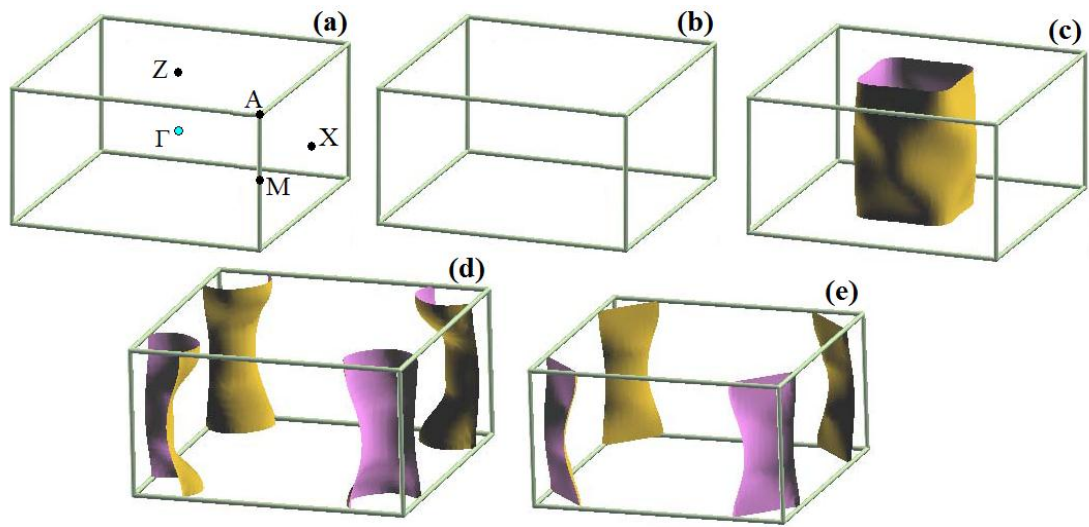

**FIG. 6** (color online) Same as Fig. 3 but for $V = 0.7\Omega_0$.

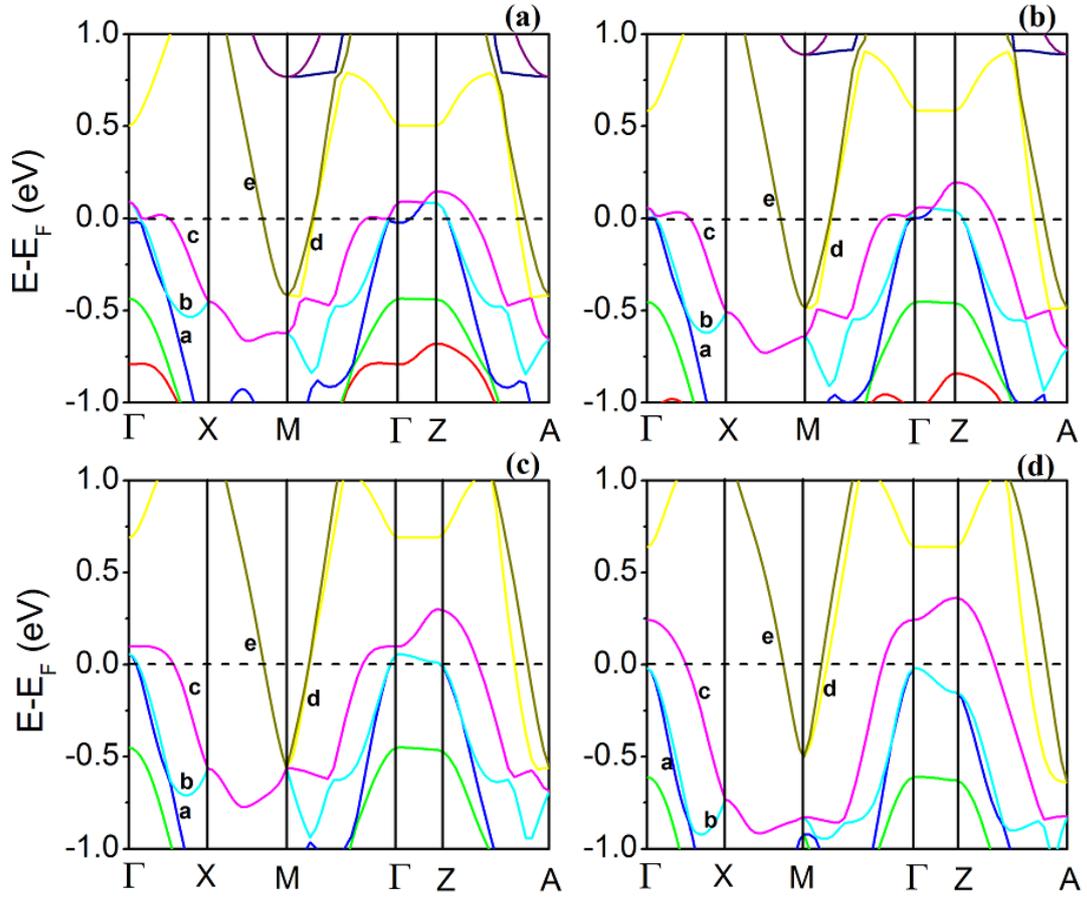

**FIG. 7** (color online) Band structures of LaOFeP at volume $V$ = **(a)** $\Omega_0$; **(b)** $0.9\Omega_0$; **(c)** $0.8\Omega_0$; **(d)** $0.7\Omega_0$. The Fermi level is set at 0 (dashed lines).

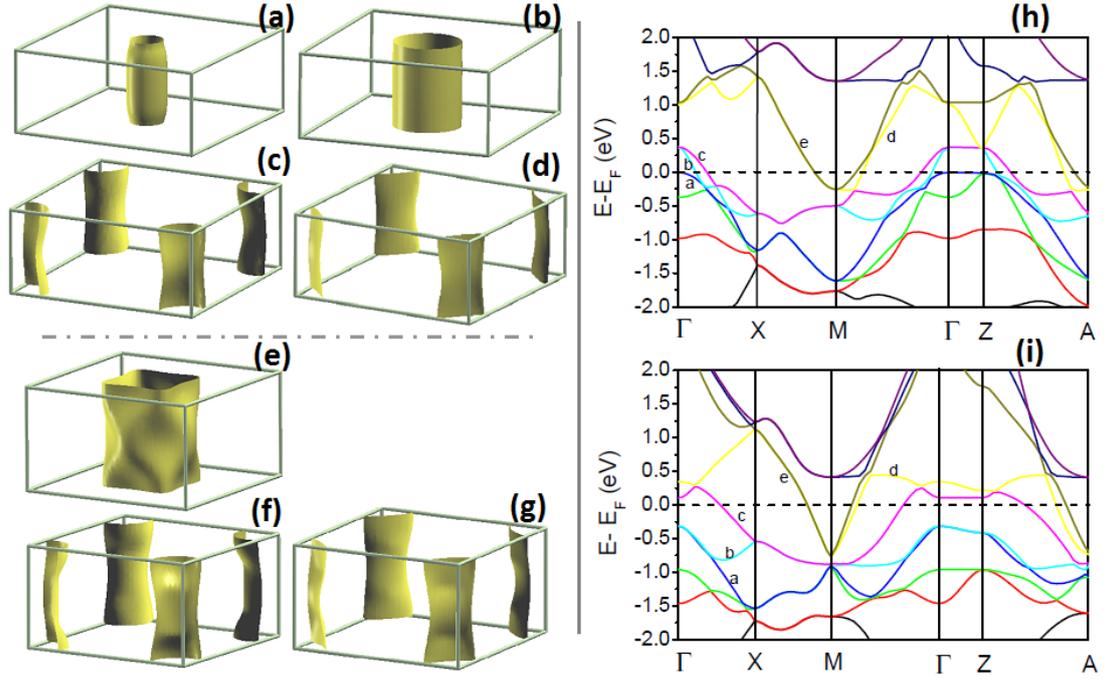

**FIG. 8** (color online) Left Panels: Fermi surface sheets of LaOFeP when compressing along *ab*-plane (panels **(a)**-**(d)**, $V \sim 0.877\Omega_0$) and along *c*-axis (panels **(e)**-**(g)**, $V \sim 0.798\Omega_0$). Right panels: band structures of *ab*-compression (panel **(h)**) and *c*-compression (panel **(i)**). The Fermi level is set at 0.

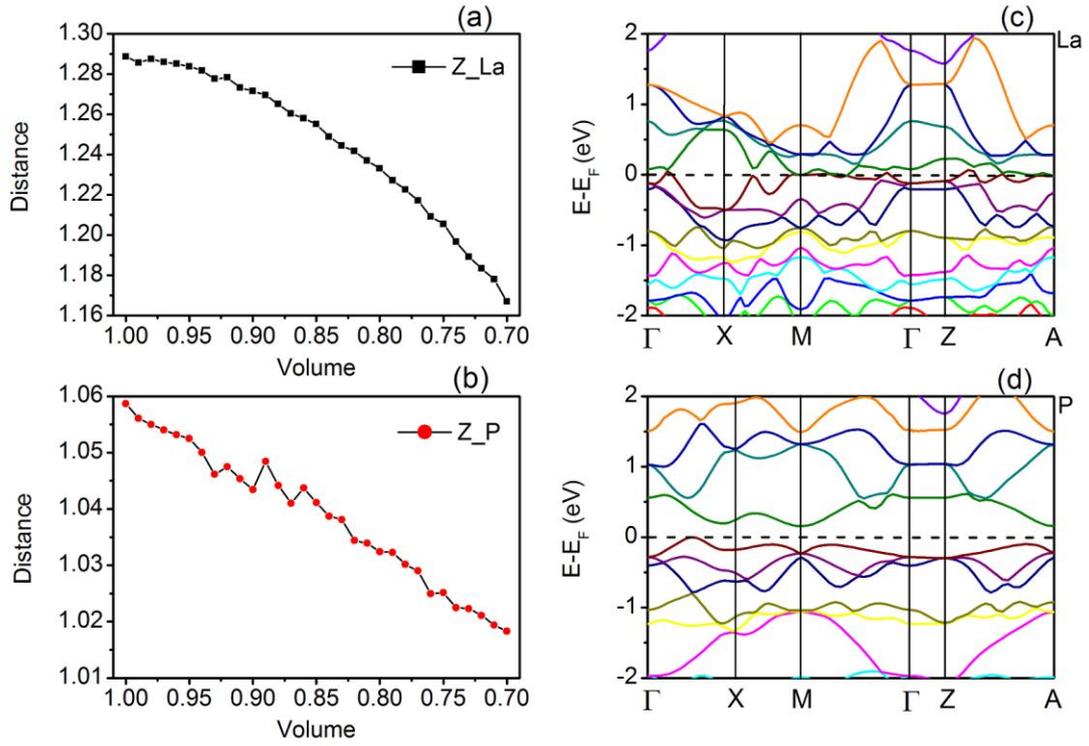

**FIG. 9** (color online) Panels **(a)-(b):** Variation of the distances of La and P atoms to the underlying O and Fe planes, respectively, as a function of normalized volume. The distances are in units of Å. Panels **(c)-(d)**: Band structures at $V = \Omega_0$, with the *z*-coordinates (the direction of *c*-axis) of La (upper panel) and P (lower panel) atoms being the values at $V = 0.7\Omega_0$.